\documentstyle[12pt,epsf]{article}

\newdimen\SaveWidth \SaveWidth=\textwidth
\newdimen\SaveHeight \SaveHeight=\textheight
\advance\SaveWidth by -\textwidth
\advance\SaveHeight by -\textheight
\divide\SaveWidth by 2
\divide\SaveHeight by 2
\advance\hoffset by \SaveWidth
\advance\voffset by \SaveHeight

\def\wtil{\widetilde}

\def\ignore#1{}
\def\beq{\begin{equation}}
\def\eeq{\end{equation}}
\def\bea{\begin{eqnarray}}
\def\eea{\end{eqnarray}}
\def\sgn{\mathop{\rm sgn}}
\def\pion{\pi}
\def\GeV{{\rm GeV}}
\def\gev{\, {\rm GeV}}
\def\TeV{{\rm TeV}}
\def\etmiss{\slashchar{E}_T}
\def\tq{{\tilde q}}
\def\tg{{\tilde g}}
\def\tchi{{\tilde\chi}}
\def\lsp{{\tilde\chi_1^0}}
\def\tell{{\tilde\ell}}
\def\ttop{{\tilde t}}
\def\tb{{\tilde b}}
\def\tG{{\tilde G}}
\def\Meff{M_{\rm eff}}
\def\mgrav{m_{3/2}}

\def\fbi{{\rm fb^{-1}}}
\def\cm{{\rm cm}}

\def\slashchar#1{\setbox0=\hbox{$#1$}           % set a box for #1
   \dimen0=\wd0                                 % and get its size
   \setbox1=\hbox{/} \dimen1=\wd1               % get size of /
   \ifdim\dimen0>\dimen1                        % #1 is bigger
      \rlap{\hbox to \dimen0{\hfil/\hfil}}      % so center / in box
      #1                                        % and print #1
   \else                                        % / is bigger
      \rlap{\hbox to \dimen1{\hfil$#1$\hfil}}   % so center #1
      /                                         % and print /
   \fi}                                         %

\def\simge{%  ``greater than about'' symbol
    \mathrel{\rlap{\raise 0.511ex
        \hbox{$>$}}{\lower 0.511ex \hbox{$\sim$}}}}
\def\simle{%  ``less than about'' symbol
    \mathrel{\rlap{\raise 0.511ex
        \hbox{$<$}}{\lower 0.511ex \hbox{$\sim$}}}}

\def\dofig#1#2{\epsfxsize=#1\centerline{\epsfbox{#2}}}
\def\dofigs#1#2#3{\centerline{\epsfxsize=#1\epsfbox{#2}%
   \hfil\epsfxsize=#1\epsfbox{#3}}}

\catcode`@=11
\def\citenum#1{\csname b@#1\endcsname}
\catcode`@=12

\begin{document}

\begin{flushright}
BNL-HET-00/1 \\
UCD-2000-5 \\
hep-ph/0001249
\end{flushright}
\bigskip

\centerline{\large\bf Anomaly Mediated SUSY Breaking at the LHC}
\medskip
\centerline{Frank E. Paige$^a$ and James Wells$^b$}
\centerline{$^a$\it Physics Department, Brookhaven National Laboratory, 
Upton, NY 11973}
\centerline{$^b$\it Physics Department, University of California, 
Davis, CA 95616}
\bigskip\bigskip

\begingroup
\narrower\narrower
\small
Anomaly Mediated SUSY Breaking models are reviewed. Possible signatures
at the LHC for one case of the minimal realistic model are examined.
\vskip0pt
\endgroup

\bigskip\bigskip
\noindent
{\it Introduction}
\bigskip

The signatures for SUSY at the LHC depend very much on the SUSY masses,
which presumably result from spontaneous SUSY breaking. It is not
possible to break SUSY spontaneously using just the MSSM fields; instead
one must do so in a hidden sector and then communicate the breaking
through some interaction. In supergravity models, the communication is
through gravity. In gauge mediated models it is through gauge
interactions; the gravitino is then very light and can play an important
role. Simple examples of both have been discussed previously. A third
possibility is that the hidden sector does not have the right structure
to provide masses through either mechanism; then the leading
contributions come from a combination of gravity and anomalies. This is
known as Anomaly Mediated SUSY Breaking (AMSB), and it predicts a different
pattern of masses and signatures.

\bigskip
\noindent
{\it Anomaly-Mediated Supersymmetry Breaking}
\bigskip

In the supersymmetric standard model there exist AMSB contributions
to the soft mass parameters that arise via the superconformal 
anomaly~\cite{Randall:1998uk, Giudice:1998xp}.
The effect
can be understood by recognizing several important features of
supersymmetric theories.  First, supersymmetry
breaking can be represented by a chiral superfield 
$\Phi=1+m_{3/2}\theta^2$ which also acts as a compensator
for super-Weyl transformations. Treating $\Phi$ as a spurion, one can
transform a theory into a super-conformally invariant theory.  
Even if a theory is superconformal at the outset (i.e., no
dimensionful couplings), the spurion $\Phi$ is employed since the
quantum field theory requires a regulator that implies scale 
dependence (Pauli-Villars
mass, renormalization scale in dimensional reduction, etc.).  
To preserve scale invariance the renormalization scale parameter $\mu$ in
a quantum theory then becomes $\mu/\sqrt{\Phi^\dagger\Phi}$.  It is
the dependence of the regulator on $\Phi$ that induces supersymmetry
breaking contributions to the scalars and gauginos.

The anomaly induced masses can be derived straightforwardly
for the scalar masses.  The K\" ahler kinetic terms depend on wave function
renormalization as in the following superfield operator,
\beq
 \int d^2\theta d^2\bar \theta Z_Q
\left( \frac{\mu}{\sqrt{\Phi^\dagger\Phi}}
 \right) Q^\dagger Q.
\eeq
Taylor expanding $Z$ around $\mu$ and projecting out the $FF^\dagger$ terms
yields a supersymmetry breaking mass for the scalar field $\tilde Q$:
\beq
\label{squarkmass}
m_{\tilde Q}^2 = -\frac{1}{4}\frac{d^2\ln Z_Q}{d(\ln \mu )^2} m^2_{3/2}
  = -\frac{1}{4}\left(\frac{\partial\gamma_Q}{\partial g}\beta_g
   +\frac{\partial \gamma_Q}{\partial y}\beta_y\right)m_{3/2}^2.
\eeq
Similar calculations can be done for the gauginos and the $A$ terms:
\bea
M_i & = & -\frac{g_i^2}{2}\frac{dg_i^{-2}}{d\ln \mu} m_{3/2} 
    =\frac{\beta_{g_i}}{g_i}m_{3/2}, \\
A_y & = & \frac{1}{2}\sum_a \frac{d\ln Z_{Q_a}}{d\ln\mu}m_{3/2}
   = -\frac{\beta_y}{y}m_{3/2}
\label{gauginomass}
\eea
where the sum over $a$ includes all fields associated with the Yukawa
coupling $y$ in the superpotential.

There are several important characteristics 
of the AMSB spectrum to note. First,
the equations for the supersymmetry breaking contributions are
scale invariant.  That is, the value of the soft masses at any scale is
obtained by simply plugging in the gauge couplings and Yukawa couplings 
at that scale into the above formulas.  Second, the masses are related to the
gravitino mass by a one loop suppression.  In AMSB 
$M_i\sim m_{3/2}\alpha_i /4\pi$, whereas in SUGRA $M_i\sim m_{3/2}$.
While the AMSB contributions are always present in a theory independent of
how supersymmetry breaking is accomplished, they may be highly
suppressed compared to standard hidden sector models.  Therefore, for
AMSB to be the primary source of scalar masses, one needs to assume or
arrange that supersymmetry breaking is not directly communicated from a
hidden sector.  This can be accomplished, for example, by assuming 
supersymmetry breaking on a distant brane~\cite{Randall:1998uk}.
Finally, the squared masses
of the sleptons are 
negative (tachyonic)
because $\beta_g>0$ for $U(1)$ and $SU(2)$ gauge groups.  This
problem rules out the simplest AMSB model based solely on 
eqs.~\ref{squarkmass}-\ref{gauginomass}.

Given the tachyonic slepton problem, it might seem most rational to
view AMSB as a good idea that did not quite work out.  However, there
are many reasons to reflect more carefully on AMSB.  As already mentioned
above, AMSB contributions to scalar masses are always present if
supersymmetry is broken.  Soft masses in the MSSM come for free, whereas
in all other successful theories of supersymmetry breaking a communication
mechanism must be detailed.  In particle, hidden sector models require
singlets to give the gauginos an acceptable mass.  
In AMSB, singlets are not necessary.
Also, there may be small variations on the AMSB idea that can
produce a realistic spectrum and can have important 
phenomenological consequences.  This is our motivation for writing 
this note. 

%%%%%%%%%%%
\bigskip
\noindent
{\it Two realistic minimal models of AMSB: mAMSB and DAMSB}
\bigskip

As we discussed in the introduction,
the pure AMSB model gives negative squared masses for the
sleptons, thus breaking electromagnetic gauge invariance, so some
additional contributions must be included. The simplest assumption
that solves this problem is to add
at the GUT scale
a single universal scalar
mass $m_0^2$ to all the sfermions' squared masses.
We will call this model mAMSB.
The description and many phenomenological implications
of this model are given in Refs.~\cite{GGW,Feng:1999hg}.
The parameters of the model after the usual
radiative electroweak symmetry breaking are then
$$
m_0,\quad \mgrav,\quad \tan\beta,\quad \sgn\mu=\pm .
$$
This model has been implemented in ISAJET~7.48~\cite{ISAJET}; a
pre-release version of ISAJET has been used to generate the events for
this analysis.

For this note the AMSB parameters were chosen to be
$$
m_0=200\,\GeV,\quad \mgrav=35\,\TeV,\quad \tan\beta=3,\quad \sgn\mu=+
$$
For this choice of parameters the slepton squared masses are positive at
the weak scale, but they are still negative at the GUT scale. This means
that charge and color might be broken (CCB) at high temperatures in the
early universe. However, at these high energies there are also large
finite temperature effects on the mass, which are positive (symmetry
restoration occurs at higher $T$). In fact, a large class of SUSY
models with CCB minima naturally fall into the correct SM minimum when
you carefully follow the evolution of the theory from high T to today.
If CCB minima are excluded at all scales, then the value of $m_0$ must
be substantially larger, so the sleptons must be quite heavy.

	The masses from ISAJET~7.48 for this point are listed in
Table~\ref{mtable}. The mass spectrum has some similarity to that for
SUGRA Point 5 studied previously~\cite{HPSSY,TDR}: the gluino and squark
masses are similar, and the decays $\tchi_2^0 \to \tell\ell$ and
$\tchi_2^0 \to \lsp h$ are allowed. Thus, many of the techniques
developed for Point 5 are applicable here. But there are also important
differences. In particular, the $\tchi_1^\pm$ is nearly degenerate with
the $\lsp$, not with the $\tchi_2^0$.  The mass splitting between
the lightest chargino and the lightest neutralino must be calculated
as the difference between the lightest eigenvalues
of the full one-loop neutralino and chargino mass matrices.
The mass splitting is always above $m_{\pion^\pm}$, thereby allowing
the two-body decays 
$\chi^\pm_1\to\chi^0_1+\pion^\pm$~\cite{GGW,Feng:1999fu}.  Decay lifetimes
of $\chi^\pm$ are always less than 10~cm over mAMSB parameter space, and
are often less than 1~cm.

Another unique feature of the spectrum is the near degeneracy of
the $\tilde\ell_L$ and $\tilde\ell_R$ sleptons.  The mass splitting 
is~\cite{GGW}
\bea
m^2_{\tell_L}-m^2_{\tell_R}\simeq 0.037 \left( -m^2_Z\cos 2\beta
  +M^2_2\ln \frac{m_{\tell_R}}{m_Z}\right).
\eea
There is no symmetry requiring this degeneracy, but rather it
is an astonishing accident and prediction of the mAMSB model.

It is instructive to compare the masses
from ISAJET with those calculated in Ref.~\citenum{GGW} to provide
weak-scale input to ISAJET. These masses are listed in the right hand
side of Table~\ref{mtable}. Since the agreement is clearly adequate for
the purposes of the present study, no attempt has been made to
understand or resolve the differences. It is clear, however, that if
SUSY is discovered at the LHC and if masses or combinations of masses
are measured with the expected precision, then more work is needed
to compare the LHC results with theoretical models in a sufficiently
reliable way.

\begin{table}[t]
\caption{Masses of the SUSY particles, in GeV, for the mAMSB model with
$m_0=200\,\GeV$, $m_{3/2}=35\,\TeV$, $\tan\beta=3$, and $\sgn\mu=+$ from
ISAJET (left side) and from Ref.~\citenum{GGW} (right side) using the
ISAJET sign conventions. \label{mtable}}
\begin{center}
\begin{tabular}{cccc|cccc} \hline \hline
Sparticle  & mass \qquad &Sparticle & mass &
Sparticle  & mass \qquad &Sparticle & mass \\ \hline
$\wtil g$  		& 815 	& 			&	&
$\wtil g$  		& 852 	& 			&	\\
$\wtil \chi_1^\pm$	& 101 	& $\wtil \chi_2^\pm$	& 658	&
$\wtil \chi_1^\pm$	& 98 	& $\wtil \chi_2^\pm$	& 535	\\
$\wtil \chi_1^0$	& 101	& $\wtil \chi_2^0$	& 322	&
$\wtil \chi_1^0$	& 98	& $\wtil \chi_2^0$	& 316	\\
$\wtil \chi_3^0$	& 652	& $\wtil \chi_4^0$	& 657	&
$\wtil \chi_3^0$	& 529	& $\wtil \chi_4^0$	& 534	\\
$\wtil u_L$		& 754	& $\wtil u_R$		& 758	&
$\wtil u_L$		& 760	& $\wtil u_R$		& 814	\\
$\wtil d_L$		& 757	& $\wtil d_R$		& 763	&
$\wtil d_L$		& 764	& $\wtil d_R$		& 819	\\
$\wtil t_1$		& 516	& $\wtil t_2$		& 745	&
$\wtil t_1$		& 647	& $\wtil t_2$		& 778	\\
$\wtil b_1$		& 670	& $\wtil b_2$		& 763	&
$\wtil b_1$		& 740	& $\wtil b_2$		& 819	\\
$\wtil e_L$		& 155	& $\wtil e_R$		& 153	&
$\wtil e_L$		& 161	& $\wtil e_R$		& 159	\\
$\wtil \nu_e$		& 137 	& $\wtil \nu_\tau$	& 137	&
$\wtil \nu_e$		& 144 	& $\wtil \nu_\tau$	& 144	\\
$\wtil \tau_1$		& 140	& $\wtil \tau_2$	& 166	&
$\wtil \tau_1$		& 152	& $\wtil \tau_2$	& 167	\\
$h^0$			& 107	& $H^0$			& 699	&
$h^0$			& 98	& $H^0$			& 572	\\
$A^0$			& 697	& $H^\pm$		& 701	&
$A^0$			& 569	& $H^\pm$		& 575	\\
\hline \hline
\end{tabular}
\end{center}
\end{table}

Another variation on AMSB is deflected AMSB (DAMSB).  The idea is
based on Ref.~\cite{Pomarol:1999ie} who demonstrated that realistic
sparticle spectrums with non-tachyonic sleptons can be induced if
a light {\it modulus} field $X$ (SM singlet) is coupled to heavy, non-singlet 
vector-like messenger fields $\Psi_i$ and $\bar\Psi_i$:
$$
W_{\rm mess} = \lambda_\Psi X\Psi_i\bar\Psi_i.
$$
To ensure gauge coupling unification we identify $\Psi_i$ and $\bar\Psi_i$
as $5+\bar 5$ representations of $SU(5)$.
When the messengers are integrated out at some
scale $M_0$, the beta
functions do not match the AMSB masses, and the masses are 
deflected from the AMSB renormalization group trajectory.  The subsequent
evolution of the masses below $M_0$ induces positive mass squared for
the sleptons, and a reasonable spectrum can result.  Although there
may be additional significant parameters associated with the generation
of the $\mu$ and $B_\mu$ term in the model, we assume for this discussion
that they do not affect the spectra of the MSSM fields.  The values of
$\mu$ and $B_\mu$ are then obtained by requiring that the conditions
for EWSB work out properly.

The parameters of DAMSB are
$$
m_{3/2},\quad n,\quad M_0,\quad \tan\beta,\quad \sgn\mu=\pm
$$
where $n$ is the number of $5+\bar 5$ messenger multiplets, and $M_0$ is
the scale at which the messengers are integrated out.
Practically, the spectrum is obtained by imposing the boundary conditions
at $M_0$, and then using SUSY soft mass renormalization group equations to
evolve these masses down to the weak scale.  Expressions for
the boundary conditions
can be found in Refs.~\cite{Pomarol:1999ie,rsw}, and details on
how to generate the low-energy spectrum are given in~\citenum{rsw}.
The resulting spectrum of superpartners
is substantially different from that of mAMSB.  The most characteristic
feature of the DAMSB spectrum is the near proximity of all superpartner
masses.  In Table~\ref{DAMSB} we show the
spectrum of a model with $n=5$, $M_0=10^{15}\gev$, and $\tan\beta =4$
as given in~\cite{rsw}.
The LSP is the lightest neutralino, which is a Higgsino.  (Actually,
the LSP is the fermionic component of the modulus $X$, but the decay
of $\chi_1^0$ to it is much greater than collider time scales.)
All the gauginos and squarks are between $300\gev$
and $500\gev$, while the sleptons and higgsinos are a bit
lighter ($\sim 150\gev$ to $\sim 250\gev$) 
in this case.  

\begin{table}[t]
\caption{Masses of the SUSY particles, in GeV, for the DAMSB model with
$n=5$, $M_0=10^{15}\, \GeV$, and $\tan\beta=4$ from Ref.~\citenum{rsw}.
\label{DAMSB}}
\begin{center}
\begin{tabular}{cccc} \hline \hline
Sparticle  & mass \qquad &Sparticle & mass \\ \hline
$\wtil g$               & 500   &                       &       \\
$\wtil \chi_1^\pm$      & 145   & $\wtil \chi_2^\pm$    & 481   \\
{$\wtil \chi_1^0$}  & 136   & $\wtil \chi_2^0$      & 152   \\
$\wtil \chi_3^0$        & 462   & $\wtil \chi_4^0$      & 483   \\
$\wtil u_L$             & 432   & $\wtil u_R$           & 384   \\
$\wtil d_L$             & 439   & $\wtil d_R$           & 371   \\
$\wtil t_1$             & 306   & $\wtil t_2$           & 454   \\
$\wtil b_1$             & 371   & $\wtil b_2$           & 406   \\
$\wtil e_L$             & 257   & $\wtil e_R$           & 190   \\
$\wtil \nu_e$           & 246   & $\wtil \nu_\tau$      & 246   \\
$\wtil \tau_1$  & 190   & $\wtil \tau_2$        & 257   \\
$h^0$                   & 98    & $H^0$                 & 297   \\
$A^0$                   & 293   & $H^\pm$               & 303   \\
\hline \hline
\end{tabular}
\end{center}
\end{table}

In summary, we have outlined
two interesting directions to pursue in modifying
AMSB to make a realistic spectrum.  The first direction we call
mAMSB, and is constructed by adding a common scalar mass to the
sfermions at the GUT scale to solve the negative squared slepton mass
problem of pure AMSB.  The other direction that we outlined is deflected
anomaly mediation that is based on throwing the scalar masses off the
pure AMSB renormalization group trajectory by integrating out heavy
messenger states coupled to a modulus.  The spectra of the two approaches
are significantly different, and we should expect the LHC signatures
to be different as well.  In this note, we study the mAMSB carefully
in a few observables to demonstrate how it is distinctive from
other, standard approaches to supersymmetry breaking, such as mSUGRA
and GMSB.

%%%%%%%%%%%%%%%%%%%%%%%%%%%%%%%%%%%%%%%%%%%
\bigskip
\noindent
{\it LHC studies of the example mAMSB model point}
\bigskip

\begin{figure}[t]
\dofigs{3in}{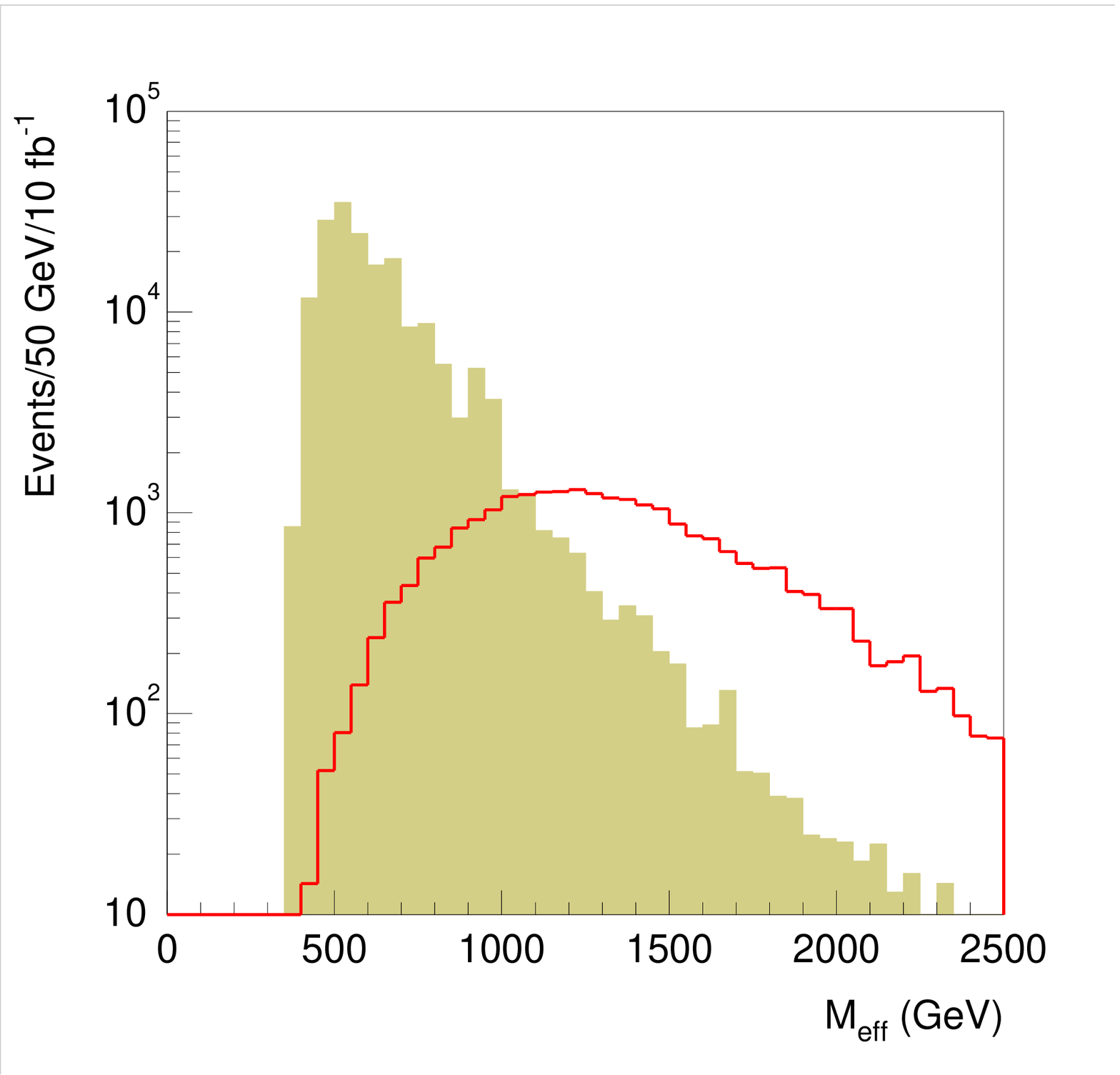}{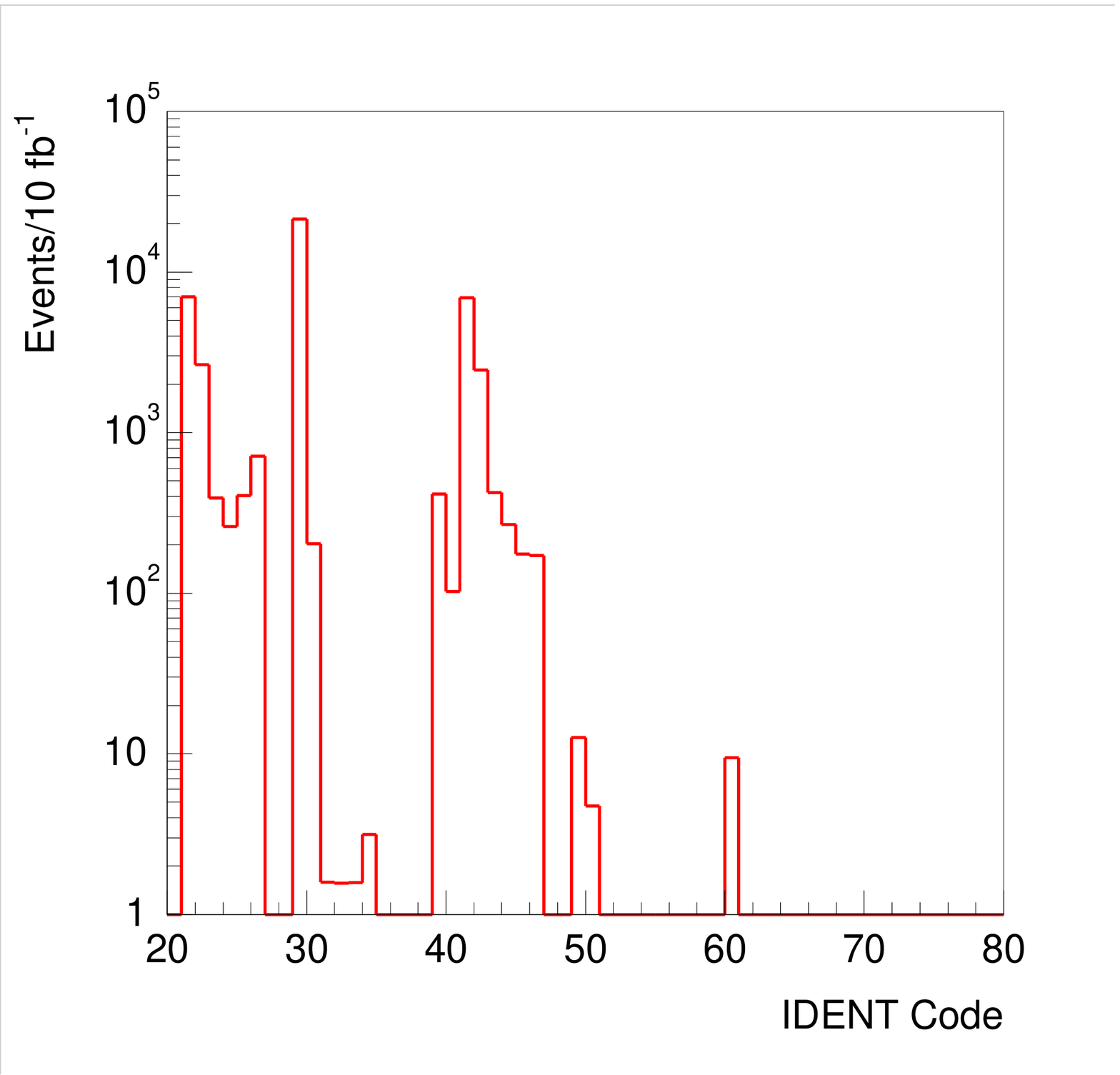}
\caption{Left: Effective mass distribution for signal (curve) and
Standard Model background (shaded). Right: ISAJET IDENT codes for all
produced particles contributing to $\Meff$ distribution after cuts. The
dominant contributions are $\tq_L$ (21--26), $\tg$ (29), and $\tq_R$
(41--46). \label{a1hmeff}}
\end{figure}

We now turn to a study of the example mAMSB spectra presented
in Table~1.
A sample of $10^5$ signal events was generated; since the total
signal cross section is $16\,{\rm nb}$, this corresponds to an
integrated LHC luminosity of $6\,\fbi$. All distributions shown in this
note are normalized to $10\,\fbi$, corresponding to one year at low
luminosity at the LHC. Events were selected by requiring
\begin{itemize}
\item At least four jets with $p_T>100,50,50,50\,\GeV$;
\item $\etmiss > \min(100\,\GeV,0.2\Meff)$;
\item Transverse sphericity $S_T>0.2$;
\item $\Meff > 600\,\GeV$;
\end{itemize}
where the ``effective mass'' $\Meff$ is given by the scalar sum of the
missing $E_T$ and the $p_T$'s of the four hardest jets,
$$
\Meff = \etmiss + p_{T,1} + p_{T,2} + p_{T,3} + p_{T,4} .
$$
Standard model backgrounds from gluon and light quark jets, $t\bar t$,
$W+{\rm jets}$, $Z + {\rm jets}$, and $WW$ have also been generated,
generally with much less equivalent luminosity. The $\Meff$
distributions for the signal and the sum of all backgrounds with all
except the last cut are shown in Figure~\ref{a1hmeff}. The ISAJET IDENT
codes for the SUSY events contributing to this plot are also shown. It
is clear from this plot that the Standard Model backgrounds are small
with these cuts, as would be expected from previous 
studies~\cite{HPSSY,TDR}.

\begin{figure}[t]
\dofigs{3in}{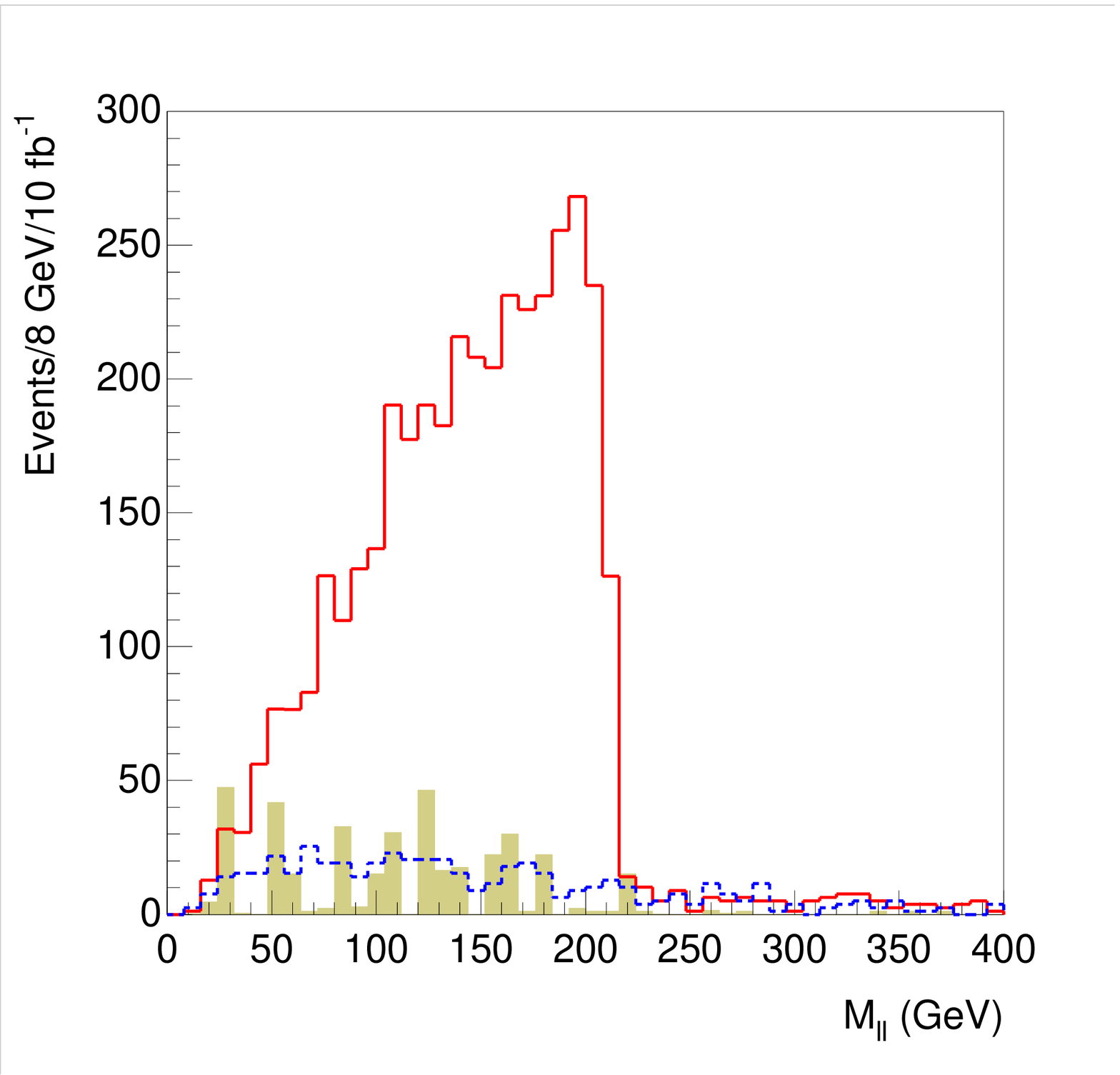}{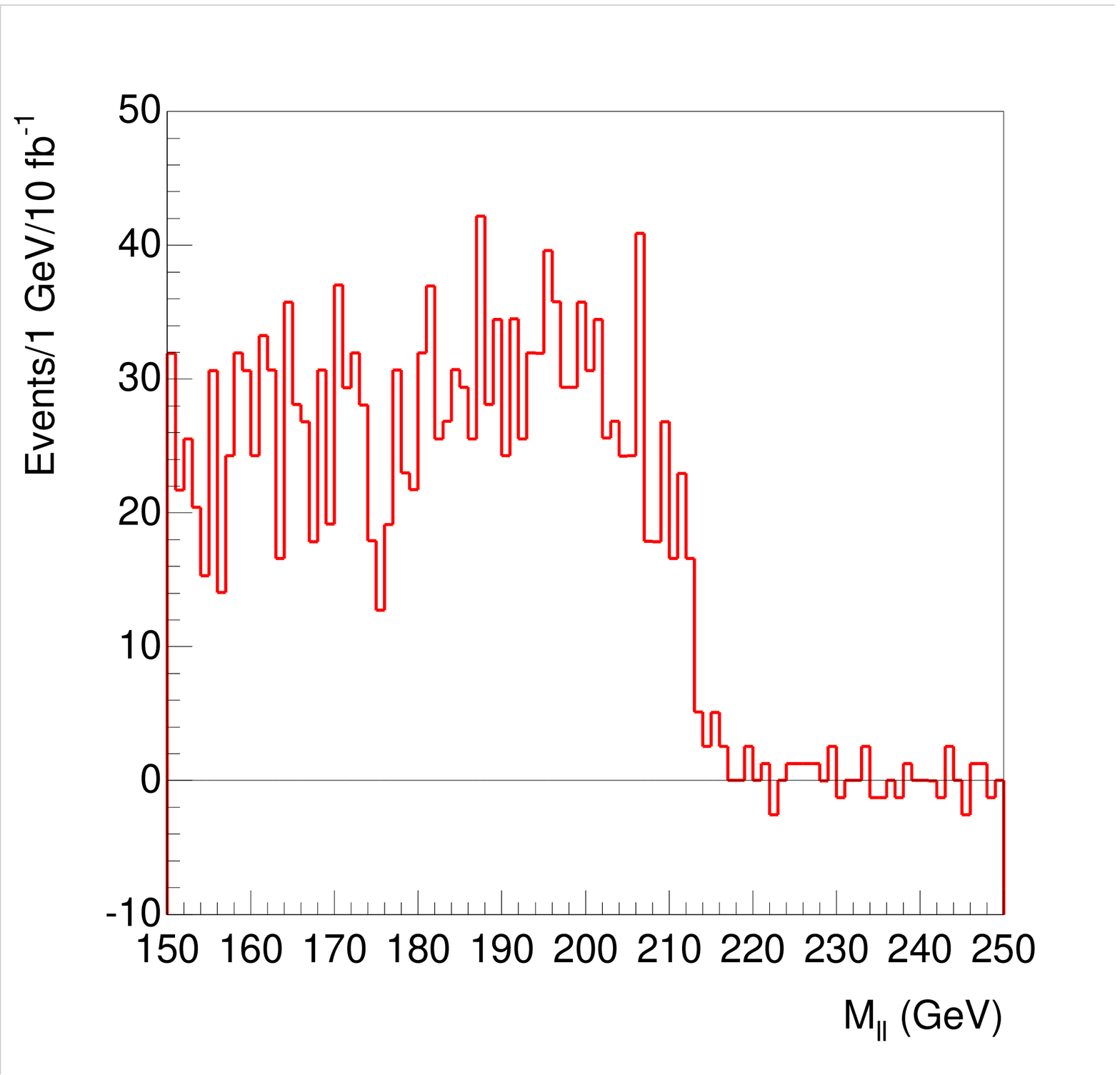}
\caption{Mass distribution for opposite sign dileptons. Left:
Distributions for same flavor signal (solid curve), opposite flavor
signal (dashed curve), and Standard Model same flavor background
(shaded). Right:  $e^+e^- + \mu^+\mu^- - e^\pm\mu^\mp$ distribution for
signal on a finer scale. \label{a1hmll}}
\end{figure}

	The mass distribution for $\ell^+\ell^-$ pairs with the same and
opposite flavor is shown in Figure~\ref{a1hmll}. The opposite-flavor
distribution is small, and there is a clear endpoint in the same-flavor
distribution at
$$
M_{\ell\ell}^{\rm max} = \sqrt{(M_{\tchi_2^0}^2 - M_{\tell}^2)
(M_{\tell}^2 - M_\lsp^2)\over M_{\tell}^2} = 213.6, 215.3\,\GeV
$$
corresponding to the endpoints for the decays $\tchi_2^0 \to
\tell_{L,R}^\pm \ell^\mp \to \lsp \ell^+\ell^-$. This is similar to what
is seen in SUGRA Point 5, but in that case only one slepton contributes.
It is clear from the $e^+e^- + \mu^+\mu^- - e^\pm\mu^\mp$ dilepton
distribution with finer bins shown in the same figure that the endpoints
for $\tell_R$ and $\tell_L$ cannot be resolved with the expected ATLAS
dilepton mass resolution. More work is needed to see if the presence of
two different endpoints could be inferred from the shape of the edge of
the dilepton distribution.

\begin{figure}[t]
\dofigs{3in}{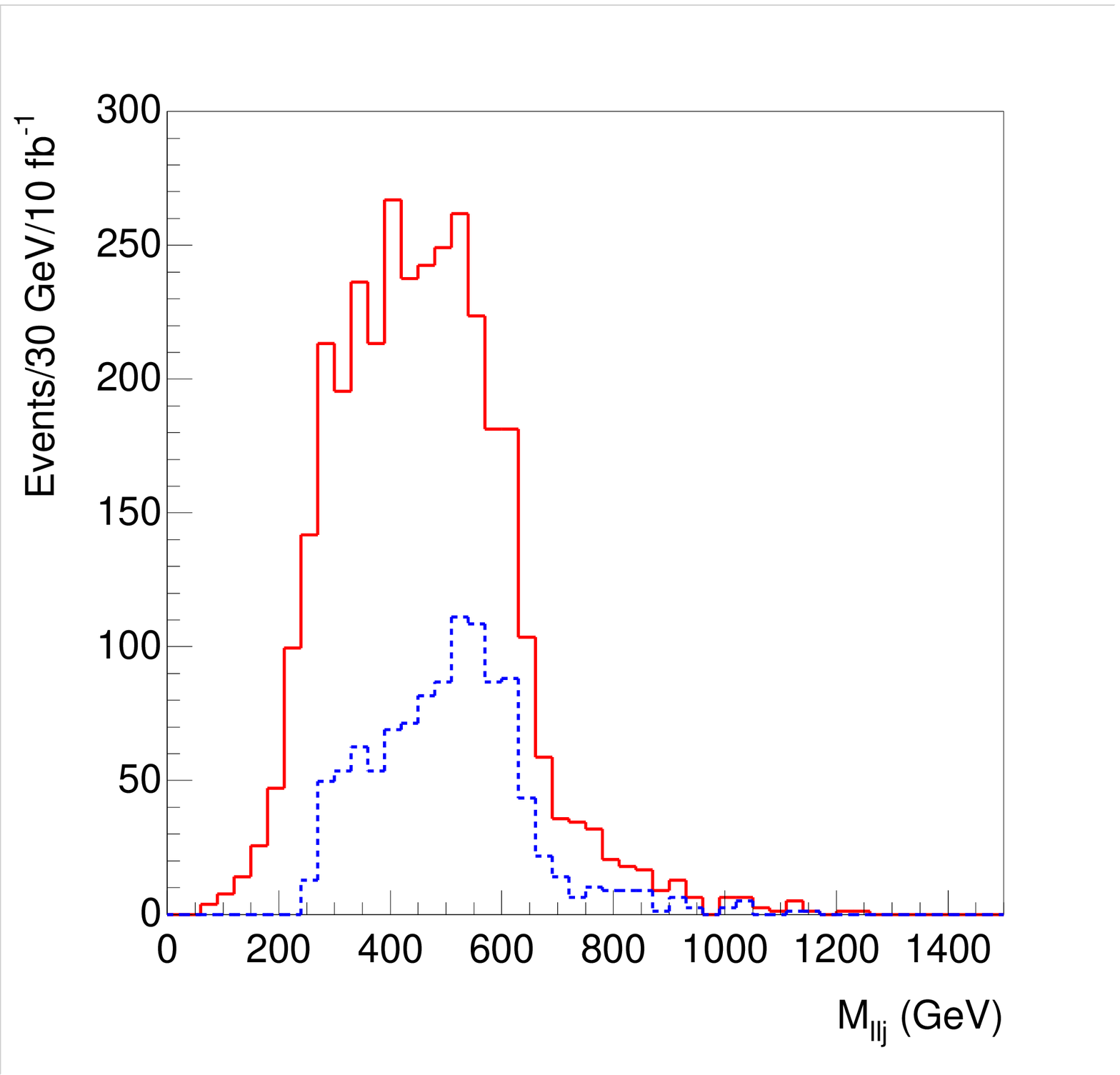}{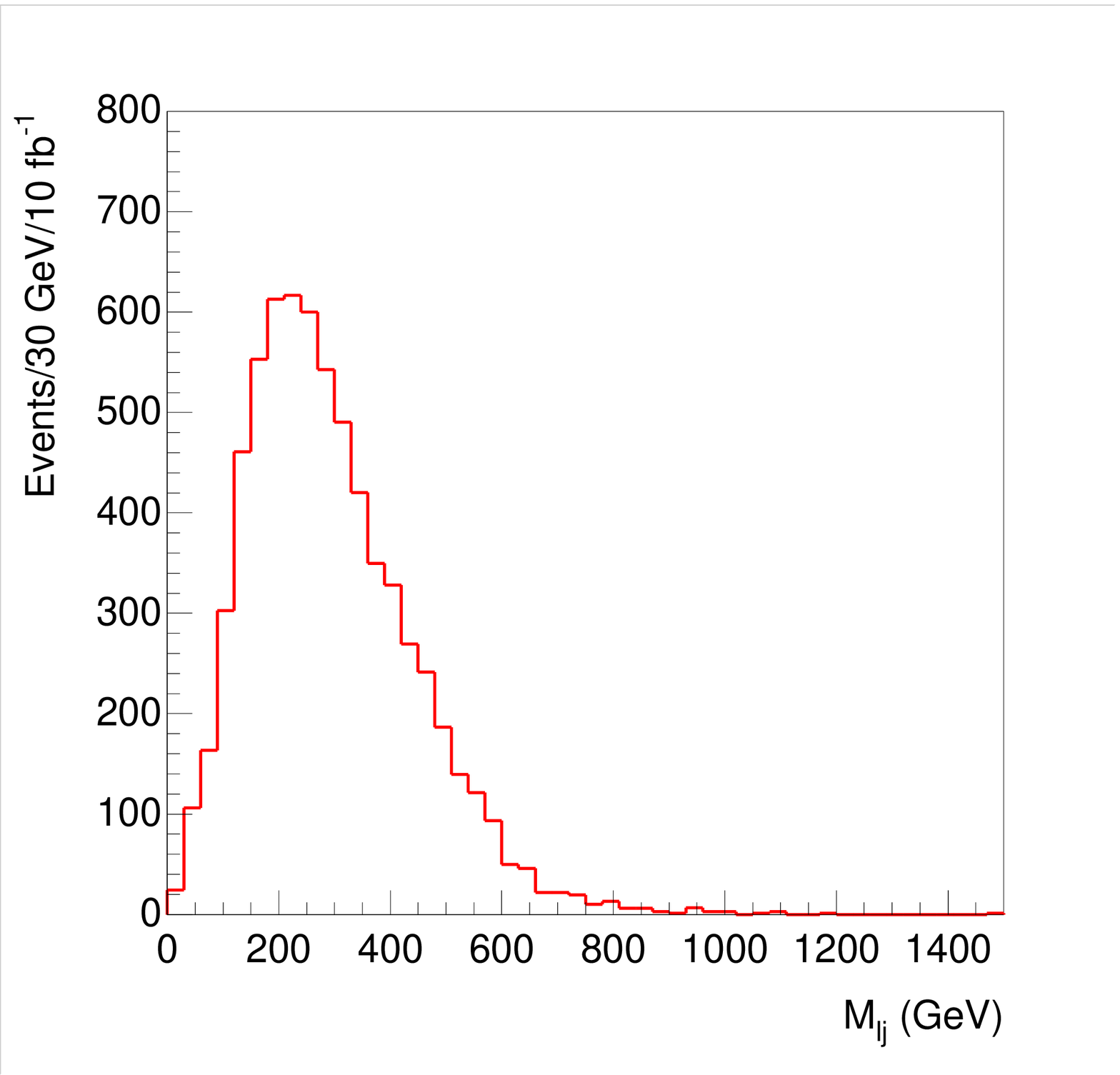}
\caption{Mass distribution for $e^+e^- + \mu^+\mu^- - e^\pm\mu^\mp$
events combined with one of the two hardest jets. Left: $\ell^+\ell^-j$
mass distribution (solid) and same with $M_{\ell\ell}>175\,\GeV$
(dashed).  Right: $\ell^\pm j$ mass distribution for the one of the two
hardest jets that gives the smaller $\ell\ell j$ mass.
\label{a1hmllj}}
\end{figure}

\begin{figure}[t]
\dofigs{3in}{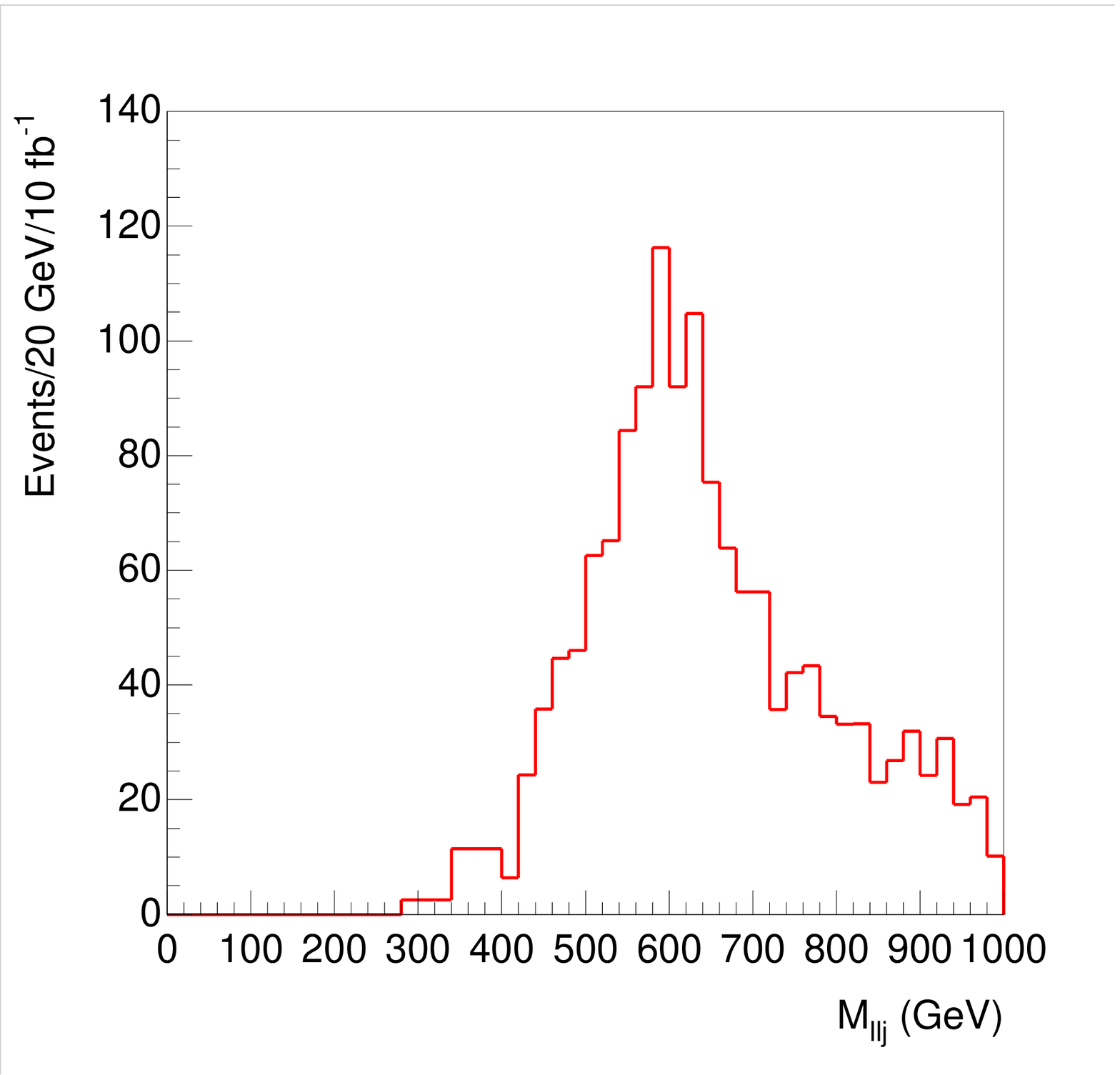}{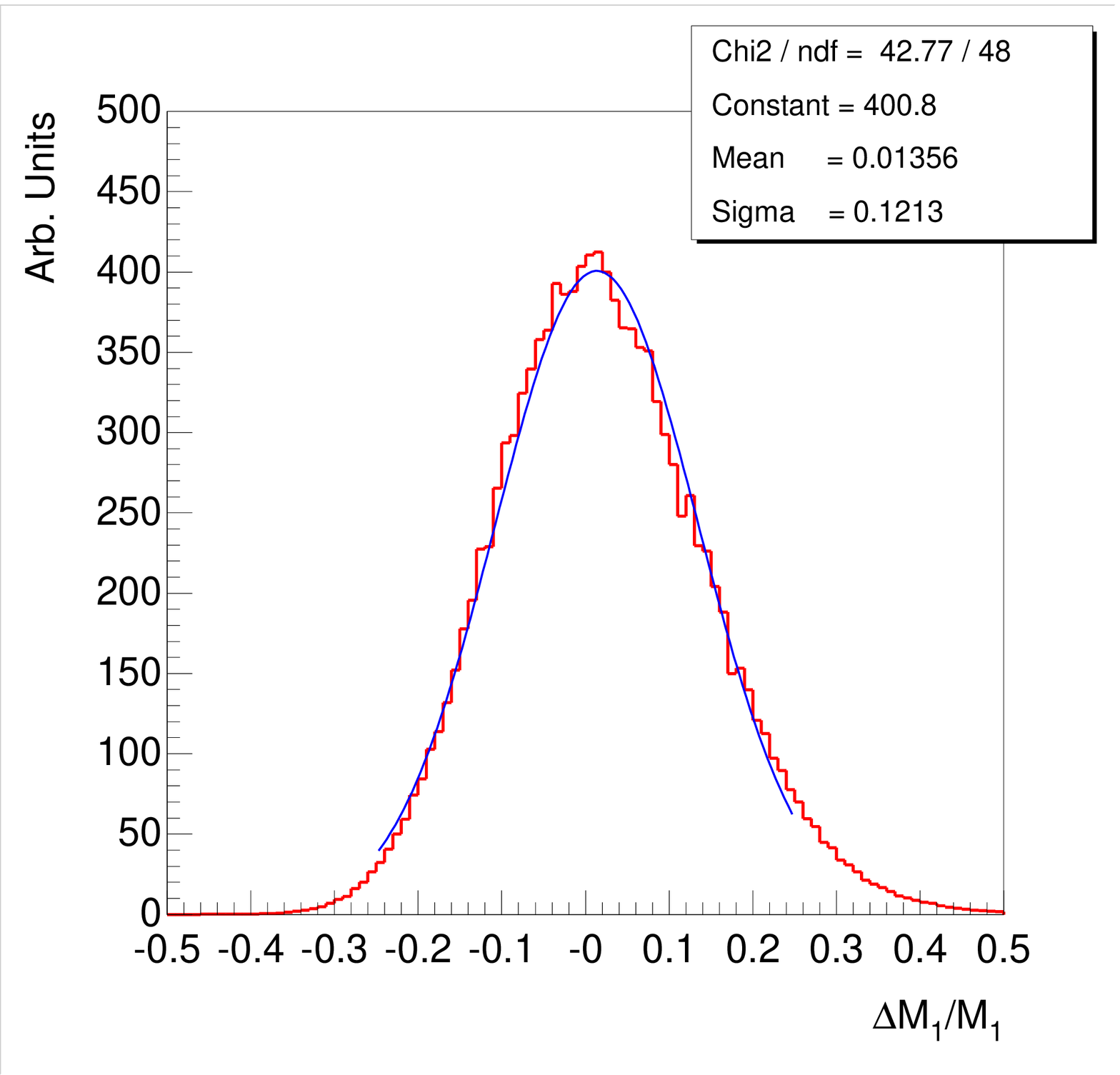}
\caption{Left: Lower edge from larger $\ell\ell j$ mass combining $e^+e^- +
\mu^+\mu^- - e^\pm\mu^\mp$ with one of the two hardest jets. Right:
Model-independent fit for $\lsp$ mass.\label{a1hmlljbk}}
\end{figure}

	Since the main source for $\tchi_2^0$ is $\tq_R \to \tchi_2^0 q$,
information on the squark masses can be obtained by combining the
leptons from $\tchi_2^0 \to \tell \ell$ decays with one of the two
hardest jets in the event, since the hardest jets are generally products
of the squark decays. Figure~\ref{a1hmllj} shows the distribution for
the smaller of the two $\ell^+\ell^- j$ masses formed with the two
leptons and each of the two hardest jets in the event. The dashed curve
in this figure shows the same distribution for $M_{\ell\ell}>175\,\GeV$,
for which the backgrounds are smaller. Both distributions should have
endpoints at the kinematic limit for $\tq_R \to \tchi_2^0 \to \tell \ell
\to \lsp \ell \ell$,
$$
\left[{(M^2_{\tq_R} - M^2_{\tchi_2^0})(M^2_{\tchi_2^0} -M^2_{\lsp}) \over
M^2_{\tchi_2^0}}\right]^{1/2} = 652.9\,\GeV .
$$
Figure~\ref{a1hmllj} also shows the $\ell^\pm j$ mass distribution
formed with each of the two leptons combined with the jet that gives the
smaller of the two $\ell\ell j$ masses. This should have a 3-body
endpoint at
$$
\left[{(M^2_{\tq_R} - M^2_{\tchi_2^0})(M^2_{\tchi_2^0} -M^2_{\tell}) \over
M^2_{\tchi_2^0}}\right]^{1/2} = 605.4\,\GeV .
$$
The branching ratio for $\tb_1 \to \tchi_2^0 b$ is very small, so the
same distributions with $b$-tagged jets contain only a handful of events
and cannot be used to determine the $\tb_1$ mass.

	The decay chain $\tq_R \to \tchi_2^0 q \to \tell_{L,R}^\pm
\ell^\mp q \to \lsp \ell^+\ell^- q$ also implies a lower limit on the
$\ell\ell q$ mass for a given limit on $z=\cos\theta^*$ or equivalently
on the $\ell\ell$ mass. For $z>0$ (or equivalently $M_{\ell\ell} >
M_{\ell\ell}^{\rm max}/\sqrt{2}$) this lower limit is
\begin{equation}
\begin{array}{lcl}
(M_{\ell\ell q}^{\rm min})^2 &=& \frac{1}{4 M_2^2 M_e^2} \times \\
&& \Biggl[-M_1^2 M_2^4 + 3 M_1^2 M_2^2 M_e^2 - M_2^4 M_e^2 - M_2^2 M_e^4
- M_1^2 M_2^2 M_q^2 - \\
&&\quad M_1^2 M_e^2 M_q^2 + 3 M_2^2 M_e^2 M_q^2 - M_e^4 M_q^2 +
(M_2^2-M_q^2)\times \\
&&\quad \sqrt{(M_1^4+M_e^4)(M_2^2 + M_e^2)^2 +
2 M_1^2 M_e^2 (M_2^4 - 6 M_2^2 M_e^2 + M_e^4)}\Biggr]\\
M_{\ell\ell q}^{\rm min}&=& 376.6\,\GeV\nonumber
\end{array}
\end{equation}
where $M_q$, $M_2$, $M_e$, and $M_1$ are the (average) squark,
$\tchi_2^0$, (average) slepton, and $\lsp$ masses. To determine this
lower edge, the larger of the two $\ell\ell j$ masses formed from two
opposite-sign leptons and one of the two hardest jets is plotted in
Figure~\ref{a1hmlljbk}. An endpoint at about the right value can clearly
be seen.

	The $\ell^+\ell^-$, $\ell^+\ell^-q$, $\ell^\pm q$, and lower
$\ell^+\ell^-q$ edges provide four constraints on the four masses
involved. Since the cross sections are similar to those for SUGRA
Point~5, we take the errors at high luminosity to be negligible on the
$\ell^+\ell^-$ edge, 1\% on the $\ell^+\ell^-q$ and $\ell^\pm q$ upper
edges, and 2\% on the $\ell^+\ell^-q$ lower edge. Random masses were
generated within $\pm50\%$ of their nominal values, and the $\chi^2$ for
the four measurements with these errors were used to determine the
probability for each set of masses. The resulting distribution for the
$\lsp$ mass, also shown in Figure~\ref{a1hmlljbk}, has a width of
$\pm11.7\%$, about the same as for Point~5; the errors for the other
masses are also comparable. Of course, the masses being measured in this
case are different: for example the squark mass is the average of the
$\tq_R$ rather than the $\tq_L$ masses.

\begin{figure}[t]
\dofig{3in}{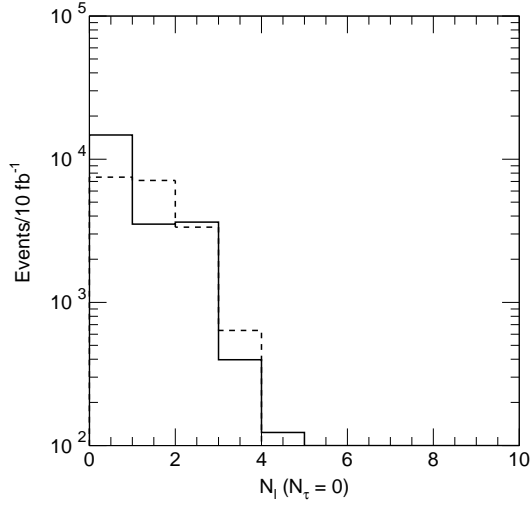}
\caption{Lepton multiplicity with a hadronic $\tau$ veto. Solid: AMSB
model. Dashed: Same but with $M_1 \leftrightarrow M_2$. \label{w1hnlep}}
\end{figure}

	The leptons from $\tchi_1^\pm \to \lsp \ell^\pm \nu$ are very
soft. This implies that the rate for events with one or three leptons or
for two leptons with opposite flavor are all suppressed.
Figure~\ref{w1hnlep} shows as a solid histogram the multiplicity of
leptons with $p_T>10\,\GeV$ and $|\eta|<2.5$ for the AMSB signal with a
veto on hadronic $\tau$ decays. The same figure shows the distribution
for a model with the same weak-scale mass parameters except that the
gaugino masses $M_1$ and $M_2$ are interchanged. This model has a wino
$\tchi_1^\pm$ approximately degenerate with the $\tchi_2^0$ rather than
with the $\lsp$. Clearly the AMSB model has a much smaller rate for
single leptons and a somewhat smaller rate for three leptons; these
rates can be used to distinguish AMSB and SUGRA-like models.

\begin{figure}[t]
\dofigs{3in}{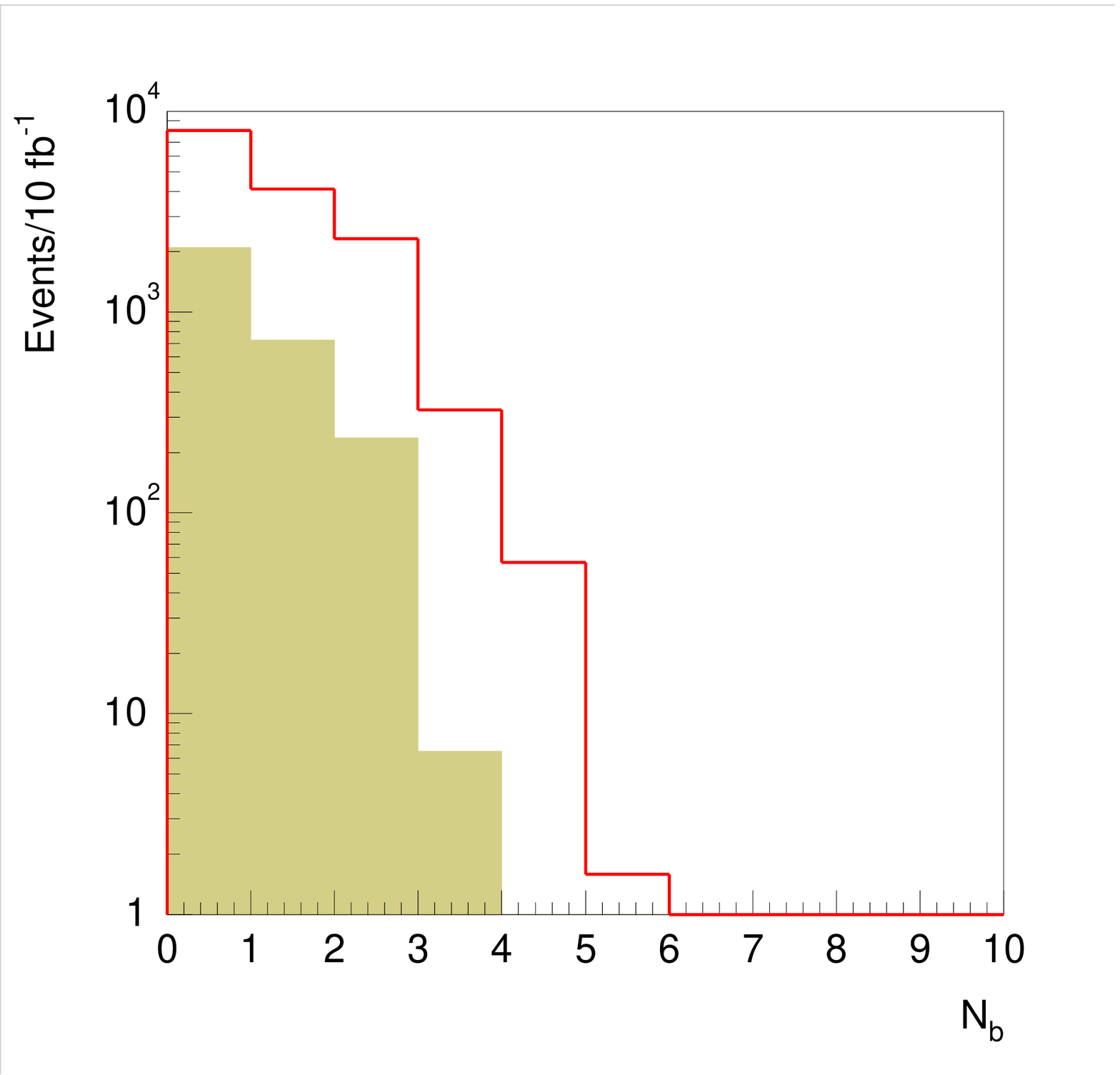}{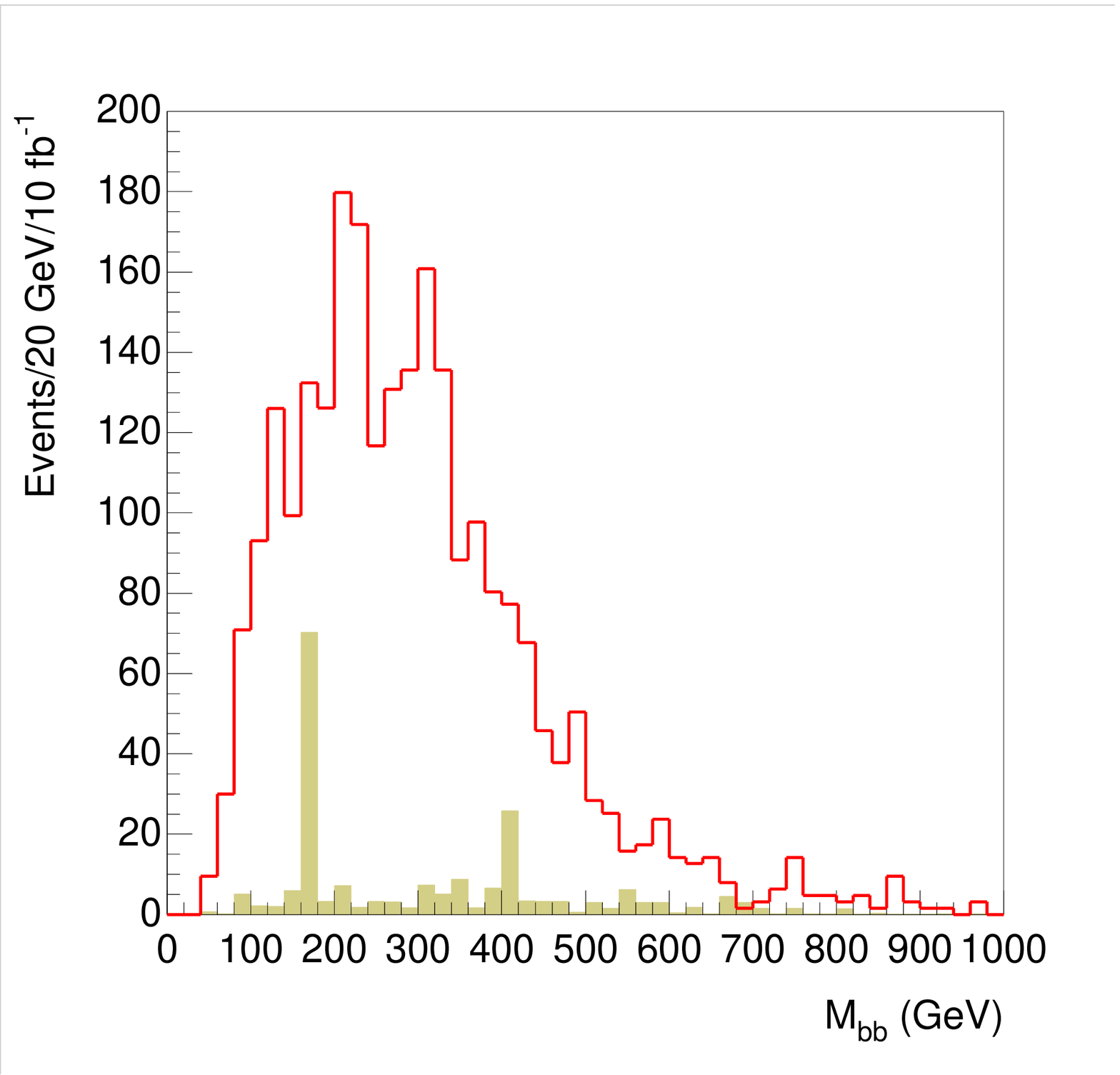}
\caption{Left: Multiplicity of $b$ jets for signal (curve) and Standard
Model background (shaded). Right: Smallest mass for pairs of $b$ jets
for signal (curve) and Standard Model background (shaded). Both plots
have $\Meff>1200\,\GeV$ in addition to the standard cuts and include a
$b$ tagging efficiency $\epsilon_b=60\%$. \label{a1hmbb}}
\end{figure}

	While the decay $\tchi_2^0 \to \lsp h$ is kinematically allowed,
the branching ratio is only about 0.03\%. Other sources of $h$ in SUSY
events are also quite small, so in contrast to SUGRA Point~5 there is no
strong $h \to b \bar b$ signal. However, there is a fairly large
branching ratio for $\tg \to \tb \bar b, \ttop \bar t$ with $\tb \to
\lsp b$, $\ttop \to \chi_1^+ b$, giving two hard $b$ jets and hence
structure in the $M_{bb}$ distribution. For this analysis $b$ jets were
tagged by assuming that any $B$ hadron with $p_{T,B}>10\,\GeV$ and
$|\eta_B|<2$ is tagged with an efficiency $\epsilon_B = 60\%$; the jet
with the smallest
$$
R = \sqrt{(\Delta\eta)^2 + (\Delta\phi)^2}
$$
was then taken to be $b$ jets. The two hardest jets generally come from
the squarks. To reconstruct $\tg \to \tb \bar b$ one of the two hardest
jets, tagged as a $b$, was combined with any remaining jet, also tagged
as a $b$. In addition to the standard multijet and $\etmiss$ cuts, a cut
$\Meff>1200\,\GeV$ was made to reduce the Standard Model background. The
resulting distributions for the $b$ jet multiplicity and for the
smallest $bb$ dijet mass are shown in Figure~\ref{a1hmbb}. The dijet
mass should have an endpoint at the kinematic limit for $\tg \to \tb_1
\bar b \to \lsp b \bar b$,
$$
M_{bb}^{\rm max} = \sqrt{(M_{\tg}^2 - M_{\tb}^2)
(M_{\tb}^2 - M_\lsp^2)\over M_{\tb}^2} = 418.7\,\GeV .
$$
While the figure is roughly consistent with this, the endpoint is not
very sharp; more work is needed to assign an error and to understand the
high mass tail. There should also be a $b\bar t$ endpoint resulting from
$\tg \to \ttop \bar t$, $\ttop \to \tchi_1^+ b$, with $M_{\tchi_1^+}
\approx M_\lsp$ and essentially invisible. Of course $m_t$ has to be
kept in the formula. This would be an apparent strong flavor violation
in gluino decays and so quite characteristic of these models.
Reconstructing the top is more complicated, so this has not yet been
studied.

\begin{figure}[t]
\dofigs{3in}{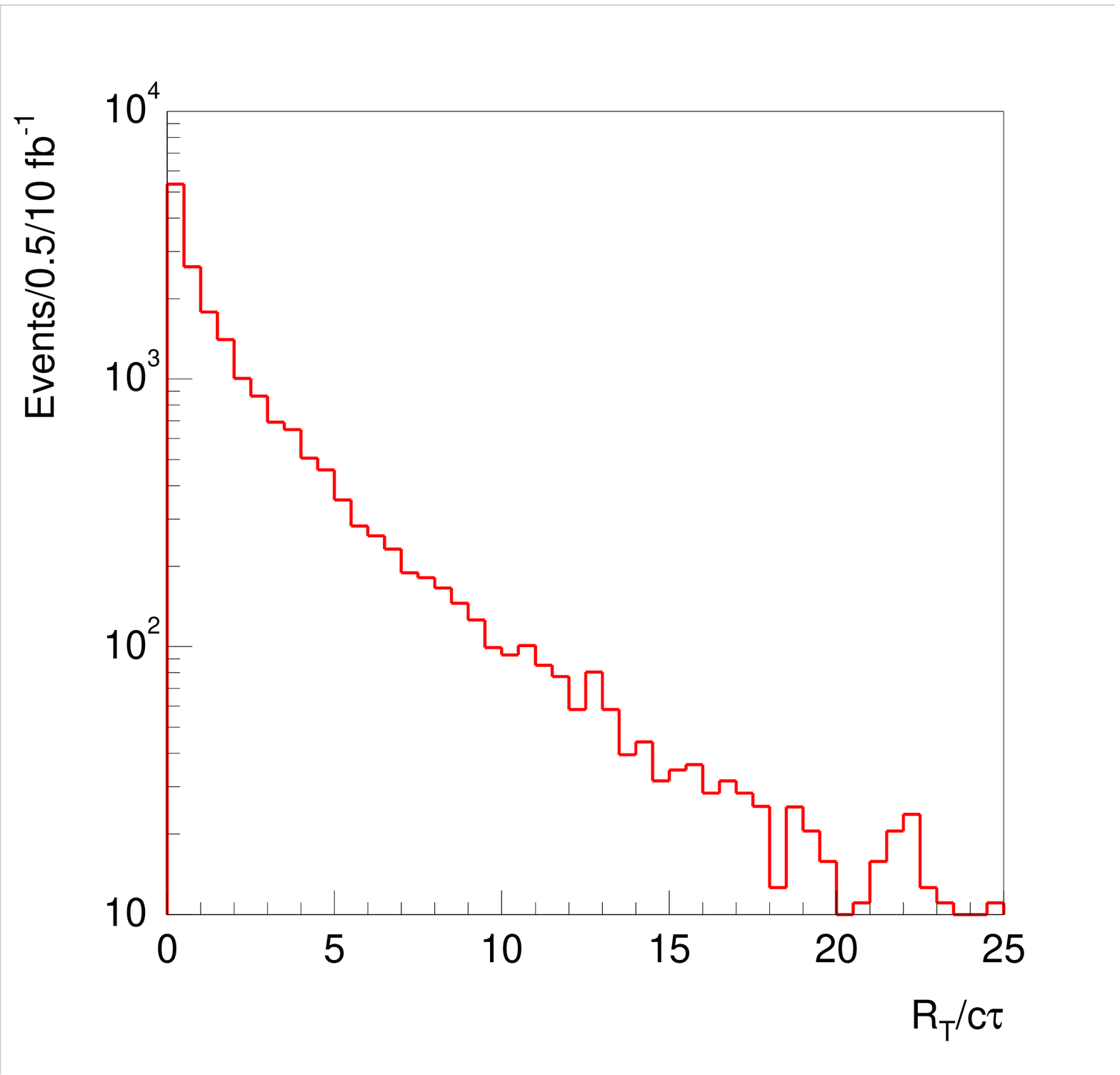}{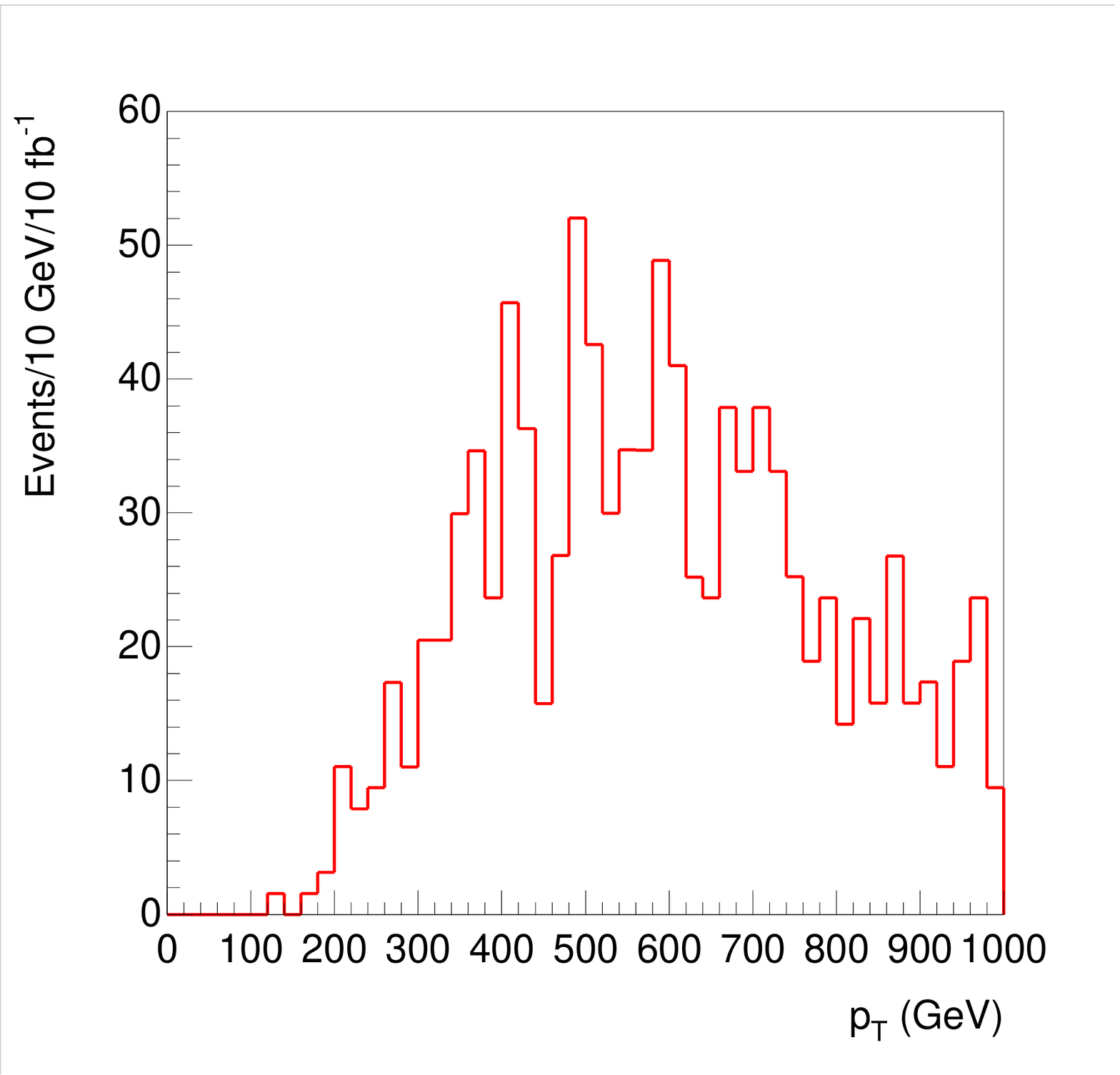}
\caption{Left: Radial track length distributions for $\tchi_1^\pm$ in
the barrel region, $|\eta|<1$. Right: $p$ distribution for $\tchi_1^\pm$
with $R_T > 10c\tau$. \label{a1hrwino}}
\end{figure}

	The splitting between the $\tchi_1^\pm$ and $\lsp$ is very small
in AMSB models. ISAJET gives a splitting of $0.189\,\GeV$ for this point
and $c\tau = 2.8\,\cm$, with the dominant decay being the two-body mode
$\tchi_1^\pm \to \lsp \pi$ via a virtual $W$. Ref.~\citenum{GGW} gives a
somewhat smaller value of $\mu$ and so a smaller splitting. The lifetime
is of course quite sensitive to the exact splitting. Since the pion or
electron is soft and so difficult to reconstruct, it seems better to
look for the tail of long-lived winos. The signature is an isolated
stiff track in a fraction of the events that ends in the tracking volume
and produces no signal in the calorimeter or muon system.
Figure~\ref{a1hrwino} shows the radial track length $R_T$ distribution
in units of $c\tau$ for winos with $|\eta|<1$ and the (generated)
momentum distribution for those with $R_T>10c\tau$. Note that the ATLAS
detector has three layers of pixels with very low occupancy at radii of
4, 11, and 14~cm and four double layers of silicon strips between 30 and
50~cm. It seems likely that the background for tracks that end after the
pixel layers would be small. 

	It is instructive to compare this signature to that for GMSB
models with an NLSP slepton. Both models predict long-lived charged
particles with $\beta<1$. In the GMSB models, two NLSP sleptons occur in
every SUSY event, and they decay into a hard $e$'s, $\mu$'s, or $\tau$'s
plus nearly massless $\tG$'s. In the AMSB models, only a fraction of the
SUSY events contain long-lived charged tracks, and these decay into a
soft pion or electron plus an invisible particle. A detailed tracking
simulation should be done for both cases.

\bigskip
\noindent
{\it Acknowledgement: }
This work was supported in part by the U.S. Department of Energy
under Contract DE-AC02-98CH10886.  We also acknowledge the support of the
Les Houches Physics Center, where part of this work was done.

%%%%%%%%%%%%%%%%%%%%%%%%%%%

\end{document}